\documentclass[journal,twoside,web]{ieeecolor2}

\usepackage[dvips]{graphics}
\usepackage{epsfig}
\usepackage{amsmath}
\usepackage{amssymb}
\usepackage{amsfonts}
\usepackage[colorlinks=false,hypertexnames=false,breaklinks=true,]{hyperref}
\usepackage{hypcap}
\usepackage{cite}
\usepackage{multirow}
\usepackage{placeins}
\usepackage{algorithm}
\usepackage{algorithmic}
\usepackage[normalem]{ulem}
\usepackage{bbm}
\usepackage{comment}
\hypersetup{pdfborder={0 0 0},pdftitle={}}
\makeatletter
\renewenvironment{figure}[1][\fps@figure]{
\edef\@tempa{\noexpand\@float{figure}[#1]}
\@tempa\capstart
}{
\end@float
}
\makeatother
\DeclareMathOperator*{\argmin}{arg\,min}

                                \usepackage{generic} \usepackage{etoolbox} \def\BibTeX{{\rm B\kern-.05em{\sc i\kern-.025em b}\kern-.08em T\kern-.1667em\lower.7ex\hbox{E}\kern-.125emX}}  \makeatletter \@ifundefined{color@begingroup} {\let\color@begingroup\relax \let\color@endgroup\relax}{} \def\fix@ieeecolor@hbox#1{ \hbox{\color@begingroup#1\color@endgroup}} \patchcmd\@makecaption{\hbox}{\fix@ieeecolor@hbox}{}{\FAILED} \patchcmd\@makecaption{\hbox}{\fix@ieeecolor@hbox}{}{\FAILED} \newcommand{\RO}{\mathrm{RO}} \newcommand{\BP}{\mathrm{BP}_{ND}} \newcommand{\Emax}{E_{\mathrm{max}}}  \pubid{\begin{minipage}[t]{\textwidth}\ \\[10pt] \centering\footnotesize{\copyright 2024 IEEE. Personal use of this material is permitted. Permission from IEEE must be obtained for all other uses, in any current or future media, including reprinting/republishing this material for advertising or promotional purposes, creating new collective works, for resale or redistribution to servers or lists, or reuse of any copyrighted component of this work in other works.} \end{minipage}}
\date{}
\title{PET mapping of receptor occupancy using joint direct parametric reconstruction}
\hypersetup{
 pdfauthor={Thibault Marin, Jinson Ouyang, Marc D. Normandin, Georges El Fakhri, Yoann Petibon},
 pdftitle={PET mapping of receptor occupancy using joint direct parametric reconstruction},
 pdfkeywords={},
 pdfsubject={},
 pdfcreator={Emacs 29.3 (Org mode 9.6.15)}, 
 pdflang={English}}
\begin{document}

\bstctlcite{IEEEexample:BSTcontrol}
\author{
  Thibault Marin, \IEEEmembership{Member, IEEE},
  Vasily Belov,
  Yanis Chemli,
  Jinsong Ouyang, \IEEEmembership{Senior Member, IEEE},
  Yassir Najmaoui,
  Georges El Fakhri, \IEEEmembership{Fellow, IEEE},
  Sridhar Duvvuri,
  Philip Iredale,
  Nicolas J. Guehl,
  Marc D. Normandin,
  Yoann Petibon
  \thanks{Research support was sponsored by Cerevel Therapeutics.  The precursor
    for MK-6884 and the standards were procured from Enigma Biomedical Group.
    Methodological developments were supported by NIH grants R01EB035093,
    R21MH121812, U01EB027003 and P41EB022544.}
  \thanks{T. Marin, V. Belov, Y. Chemli, J. Ouyang, G. El Fakhri, N. J. Guehl,
    M. D. Normandin and Y. Petibon were with the Gordon Center for Medical
    Imaging, Department of Radiology, Massachusetts General Hospital, Harvard
    Medical School.  T. Marin, Y. Chemli, J. Ouyang, G. El Fakhri, N. J. Guehl,
    M. D. Normandin are now with the Yale PET Center, Department of Radiology
    and Biomedical Imaging, Yale University, New Haven CT
    (\{thibault.marin,yanis.chemli,jinsong.ouyang,georges.elfakhri,\ %
    nicolas.guehl,marc.normandin\}@yale.edu).  V. Belov is now with Sanofi
    (vasily.belov@sanofi.com).  Y. Petibon is now with Takeda
    (yoann.petibon@takeda.com).}
  \thanks{Y. Najmaoui is with the Universit{\'e} de Sherbrooke, Department of
    Computer Engineering, QC, Canada (yassir.najmaoui@USherbrooke.ca).}
  \thanks{S. Duvvuri and P. Iredale are with Cerevel Therapeutics, Cambridge MA
    (\{sridhar.duvvuri,philip.iredale\}@cerevel.com).}
  \thanks{M. Normandin and Y. Petibon are co-corresponding authors.}}
\maketitle

\label{sec:orge183634}

\begin{abstract}
  Receptor occupancy (RO) studies using PET neuroimaging play a critical role in
  the development of drugs targeting the central nervous system (CNS).  The
  conventional approach to estimate drug receptor occupancy consists in
  estimation of binding potential changes between two PET scans (baseline and
  post-drug injection).  This estimation is typically performed separately for
  each scan by first reconstructing dynamic PET scan data before fitting a
  kinetic model to time activity curves.  This approach fails to properly model
  the noise in PET measurements, resulting in poor RO estimates, especially in
  low receptor density regions.  Objective: In this work, we evaluate a novel
  joint direct parametric reconstruction framework to directly estimate
  distributions of RO and other kinetic parameters in the brain from a pair of
  baseline and post-drug injection dynamic PET scans.  Methods: The proposed
  method combines the use of regularization on RO maps with alternating
  optimization to enable estimation of occupancy even in low binding regions.
  Results: Simulation results demonstrate the quantitative improvement of this
  method over conventional approaches in terms of accuracy and precision of
  occupancy.  The proposed method is also evaluated in preclinical in-vivo
  experiments using 11C-MK-6884 and a muscarinic acetylcholine receptor 4
  positive allosteric modulator drug, showing improved estimation of receptor
  occupancy as compared to traditional estimators.  Conclusion: The proposed
  joint direct estimation framework improves RO estimation compared to
  conventional methods, especially in intermediate to low-binding regions.
  Significance: This work could potentially facilitate the evaluation of new
  drug candidates targeting the CNS.%
\end{abstract}


\section{Introduction}
\label{sec:org207ea97}

The development of treatments for central nervous system (CNS) disorders is
impeded by high production and evaluation costs as well as high failure rates,
hindering the development of new drugs \cite{DiMasi2003}.  Therefore, there is a
critical need for methods allowing to assess new drug candidates \emph{in vivo} at
very early phases of development \cite{Wong2009}.  Thanks to its unique ability
to measure the distribution and concentration of numerous CNS targets in the
living brain, neuroimaging with positron emission tomography (PET) is a widely
used technology to screen new drug candidates, providing critical information on
drug brain penetration, target engagement (i.e. occupancy) and other
pharmacological effects.  In particular, receptor occupancy (RO) studies using
PET are now commonly used in early phases of drug development to quantify target
engagement at different dose levels and to evaluate drug efficacy
\cite{Lee2006,Khodaii2022}.

Receptor occupancy, i.e. the proportion of receptor sites occupied by the drug,
is typically measured by imaging a cohort of subjects with a pair of dynamic PET
scans, one acquired at baseline and the other after administration of the drug,
using either a radiolabeled version of the drug or a radioligand binding to the
same target.  Depending upon whether a reference region is available or not for
the radioligand at hand, different methods can be used to estimate receptor
occupancy.  For tracers for which no reference region exists, occupancy is
typically estimated using the Lassen plot \cite{Cunningham2010,Lassen1995}.  The
Lassen plot is a graphical method by which differences between baseline and
post-drug total volumes of distribution~(\(V_T\)) are plotted against
baseline \(V_T\) values across multiple brain regions, allowing the estimation of
a unique occupancy value for the whole brain by linear regression analysis.
When a reference region is available for the radiotracer, which is the focus of
the present study, specific binding can be estimated for each scan and an image
of RO can be produced by measuring relative decreases in binding potential in
each voxel between baseline and post-drug scans.  Estimation of RO is typically
achieved by first reconstructing the dynamic PET data for each scan separately,
fitting a kinetic model to the reconstructed time activity curves and
calculating RO \cite{Naganawa2016,Naganawa2019}.

Kinetic modeling using such two-step approaches (also called indirect methods)
\cite{Naganawa2016,Naganawa2019,Belov2024} results in kinetic
parameter estimates with large variance and potentially large bias, due to the
high noise level in dynamic PET reconstructions and the level of smoothing that
is often applied to PET images prior to model fitting in order to mitigate for
the noise effects.  Poor estimation performance can in part be explained by the
failure to properly model data noise during the fitting step, in which least
squares fitting is typically used but is not well suited for the noise present
in dynamic PET reconstructions.

To improve the estimation of kinetic parameters, numerous direct reconstruction
approaches have been proposed.  Direct methods formulate an end-to-end objective
function relating PET measurements to kinetic parameters directly
\cite{Kotasidis2014}.  When the kinetic model is linear (e.g. Patlak model
\cite{Angelis2011}), it can be integrated in the forward PET model and
parametric images can be estimated using the traditional maximum likelihood
expectation maximization (ML-EM) algorithm.  In the general case, optimization
transfer can be used to accelerate convergence and allow the use of arbitrary
kinetic models, linear or nonlinear \cite{Kamasak2005,Wang2010,Wang2012}.  Among
these approaches, the nested EM method has been extensively studied in the
literature \cite{Wang2013,Karakatsanis2016} and combined with other techniques
to allow joint parametric reconstruction and motion estimation \cite{Jiao2017},
partial volume correction \cite{Gong2018} or regularization with anatomical
information \cite{Loeb2015}.  Alternative solvers have also been proposed, such
as preconditioned conjugate gradient
\cite{Rakvongthai2013,Petibon2017,Petibon2020}, nested iterative coordinate
descent \cite{Kamasak2005} or the alternating direction method of multipliers
(ADMM) \cite{Gong2017}.  However, these methods only estimate kinetic parameters
from a single PET study, therefore requiring a post-processing step to calculate
RO.

In this work, we reformulate the direct reconstruction problem into a joint and
direct problem that not only estimates kinetic parameters for each scan
(baseline and post-drug) but also receptor occupancy, which is dependent on both
scans.  The joint-direct objective function is minimized using the ADMM
framework \cite{Boyd2011} and integrates a prior on the kinetic parameter maps.
This prior is composed of a variance penalty applied on the receptor occupancy
map, exploiting the fact that receptor occupancy is relatively uniform
throughout the brain~\cite{Cunningham2010} but allowing for spatial
variations \cite{deLaat2021}, and a conventional smoothness anatomical prior
\cite{Vunckx2012} applied to other model parameters.  This work elaborates on the preliminary simulation \cite{Marin2020d} and in-vivo \cite{Marin2022} results previously presented.  Here, we report a full
characterization of the proposed method in terms of bias-variance trade-off as
well as application to \emph{in-vivo} occupancy scans.  Note that joint
reconstruction priors have been proposed for longitudinal PET studies
\cite{Ellis2017,Ellis2018}, however they targeted dynamic but non-parametric PET
reconstruction problems and modeled the correlation between longitudinal scans,
whereas the current work exploits the uniformity of the receptor occupancy map,
a kinetic parameter which links the baseline and post-drug scans.  We evaluate
the proposed method both on simulation phantoms and \emph{in-vivo} nonhuman primate
data using [\textsuperscript{11}C]MK-6884 \cite{Duvvuri2021,Li2022}, a
muscarinic acetylcholine receptor 4 (M4) ligand and CVL-231, an M4 positive
allosteric modulator currently under development.

The rest of the paper is organized as follows.  Section~\ref{tmi_methods}
describes the proposed reconstruction framework, the evaluation and
implementation details.  Section~\ref{tmi_results} describes the numerical
simulation and the \emph{in-vivo} results, and section~\ref{tmi_discussions}
discusses perspectives offered by the proposed method.

\begin{figure}[htbp]
\centering
\includegraphics[trim=0.1in 0.2in 0.1in 0]{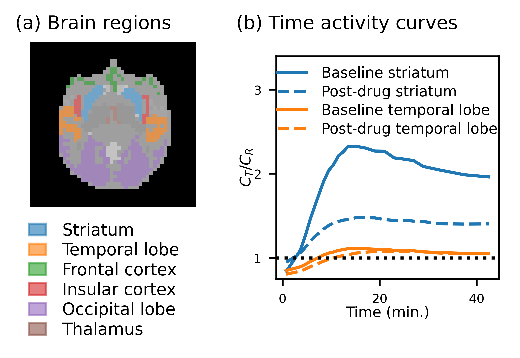}
\caption{\label{fig-phant}NHP phantom used for numerical evaluation. (a) Regions of the phantom obtained from the NMT brain atlas, (b) ratio between target $C_T$ and reference region $C_R$ time activity curves for striatum and temporal lobe regions for baseline and ``post-drug'' scans (RO=0.66).}
\end{figure}

\section{Methods}
\label{tmi_methods}
\subsection{Conventional approaches for receptor occupancy estimation}
\label{sec:org66f1ad2}

Receptor occupancy is typically estimated from two dynamic PET studies: a
baseline and a post-drug injection scan showing the blocking effect of the drug.
Conventional ``indirect'' estimation is performed following three steps.

\subsubsection{Image reconstruction}
\label{sec:orgc1e5265}

Conventional estimation methods first perform dynamic image reconstruction from
the PET projection measurements using the following objective function:
\begin{align}
\label{eq-obj-rec}
  x^{(i)*} = \arg\max_x \mathcal{L}(y^{(i)} | x) - \Psi(x),
\end{align}
where \(x^{(i)*}\) is the reconstructed dynamic PET image for scan \(i \in
\left\{B, D\right\}\) (for (B)aseline and post-(D)rug scans respectively),
\(y^{(i)}\) denotes the dynamic PET projection data for scan \(i\), \(\mathcal{L}\)
denotes the Poisson log-likelihood function and \(\Psi\) is an optional prior
term.  The forward model can be expressed as:
\begin{align}
\label{eq-fwd-rec}
  \bar{y}^{(i)} = A^{(i)}\,x^{(i)} + r^{(i)} + s^{(i)}
\end{align}
where \(\bar{y}^{(i)}\) represents expected PET measurements for image \(x^{(i)}\),
\(A^{(i)}\) is the system matrix including geometrical effects, subject motion,
attenuation coefficients and detector normalization factors, and \(r^{(i)}\),
\(s^{(i)}\) are the randoms and scatter estimates.  Equation~(\ref{eq-obj-rec})
can be solved using ordered-subset expectation-maximization (OSEM)
\cite{Hudson1994} or block sequential regularized expectation maximization
(BSREM) if a prior is included in the cost function \cite{Ahn2003}.

\subsubsection{Kinetic modeling}
\label{sec:org5238b75}

Reconstructed PET images are then used to estimate kinetic parameters using an
appropriate kinetic model \cite{Gunn2001a}.  In this work, we used the
simplified reference tissue model (SRTM) \cite{Lammertsma1996} to model time
activity curves (TAC), however the method can be generalized to other reference
region-based kinetic models (see section~\ref{tmi_discussions} for details).
Kinetic modeling can be performed either regionally or pixel-wise, which is the
focus of this work.  Under the SRTM, the instantaneous concentration in a target
region \(T\), \(C_T^{(i)}(t)\) in a single pixel for the \(i\)-th scan can be
expressed as:
\begin{IEEEeqnarray}{rCl}
  C_T^{(i)}(t)&=&R_1^{(i)} C_R^{(i)}(t) +%
  \left(k_2^{(i)} - \frac{k_2^{(i)}R_1^{(i)}}
    {1 + \BP^{(i)}}\right)\IEEEnonumber\\
  &&\quad\cdot\, C_R^{(i)}(t) \ast \exp\left(-\frac{k_2^{(i)}}
    {1 + \BP^{(i)}} t\right),
\label{eq-srtm}
\end{IEEEeqnarray}
where \(C_R^{(i)}\) is the TAC in the reference region for scan \(i\), \(R_1^{(i)}\)
is the ratio of radioligand plasma-to-tissue transport rates in target and
reference regions, \(k_2^{(i)}\) is the clearance rate, \(\BP^{(i)}\) is the binding
potential with respect to the non-displaceable radioligand uptake
\cite{Innis2007} and \(\ast\) denotes the convolution operator.  The set of
kinetic parameter maps is denoted by \(\theta^{(i)} = [\BP^{(i)}, k_2^{(i)},
R_1^{(i)}]\) and the model-predicted time-dependent activity concentration is
expressed as \(C^{(i)}(\theta^{(i)})\).  \(\theta^{(i)}\) is typically estimated
from reconstructed dynamic PET images \(x^{(i)}\) by solving a weighted
least-squares problem:
\begin{align}
\label{eq-kin-fit-basic}
  \theta^{(i)*} = \arg\min_{\theta} \frac{1}{2}
  \left\|x^{(i)} - \Gamma C^{(i)}(\theta)\right\|_W^2,
\end{align}
where \(\Gamma\) is an operator accounting for kinetic model integration over time
frames and radioactive decay and \(W\) is a diagonal weighting matrix formed based
on the number PET counts and frame duration as used in \cite{Gunn1997}.  This
fitting step can be performed independently for each pixel using different
approaches \cite{Yaqub2006}, such as the Levenberg-Marquardt algorithm, possibly
using annealing or basis functions \cite{Gunn1997}.  We focus here on the basis
function approach, which decouples the linear and nonlinear components of the
kinetic model and solves a simpler quadratic optimization problem with linear
operators for a grid of bases holding the nonlinear component.

\subsubsection{Calculation of receptor occupancy and data analysis}
\label{sec:org5e75e09}

The final step consists in combining the kinetic parameters estimated for each
scan to calculate the receptor occupancy \(\RO\), defined as the relative decrease
in \(\BP\) between scans:
\begin{equation}
\RO = 1 - \frac{\BP^{(D)}}{\BP^{(B)}}.
\end{equation}

In receptor occupancy studies, RO is typically estimated from pairs of scans for
multiple injected drug concentrations and subjects.  \(\RO\) estimates are then
plotted against drug concentration and a Hill model is fitted to the data
\cite{MacDougall2006}:
\begin{equation}
\label{eq-hill}
\RO(c) = \frac{\Emax}{1 + \frac{ED_{50}}{c}},
\end{equation}
where \(\Emax\) is the maximum drug occupancy, \(ED_{50}\) is the drug dose
resulting in a 50\% occupancy and \(c\) is the dose.  In this work, we focus on the estimation of $ED_{50}$ and $\Emax$ is known and assumed to be 100\%.

\subsection{Proposed algorithm for receptor occupancy estimation}
\label{tmi_method_admm}
The proposed method estimates RO directly from the PET projection measurements
and jointly from baseline and post-drug scans.  The estimation is driven by an
objective function connecting kinetic parameters to raw PET data:
\begin{align}
\label{eq-obj-direct}
  \theta^{(i)*} = \arg\min_{\theta} -\mathcal{L}(y^{(i)} | \theta) +
  \Phi(\theta),
\end{align}
where \(\mathcal{L}\) is a Poisson log-likelihood function, \(\Phi\) is a penalty
applied to kinetic parameter maps and the imaging model is:
\begin{align}
  \bar{y}^{(i)} =& A^{(i)}\,\Gamma\,C^{(i)}(\theta^{(i)}) +
                   r^{(i)} + s^{(i)},
\end{align}
where \(C^{(i)}\) is the kinetic model operator calculating time activity curves
from kinetic parameters.  We combine direct parametric reconstruction with joint
estimation of baseline and post-drug scans to allow estimation of receptor
occupancy directly from projection measurements.  In this framework, the
objective function can be expressed jointly for both scans:
\begin{align}
\label{eq-obj-joint-direct}
  \theta^* = \arg\min_{\theta} -\mathcal{L}(y | \theta) +
  R(\theta),
\end{align}
where \(y=[y^{(B)}, y^{(D)}]^\top\) denotes the concatenation of PET data from
both scans, \(\theta\) is the set of joint kinetic parameters, given by \(\theta =
[\BP^{(B)}, k_2^{(B)}, R_1^{(B)}, \RO, k_2^{(D)}, R_1^{(D)}]^\top\) and \(R\) is a
penalty applied on \(\theta\) described below.  The corresponding imaging model is
given by:
\begin{align}
  \bar{y} =& A\,\Gamma\,C(\theta) + r + s,
\end{align}
where \(\bar{y} = [\bar{y}^{(B)}, \bar{y}^{(D)}]^\top]\), \(A\) is the concatenation
of the baseline and post-drug system matrices, \(r = [r^{(B)}, r^{(D)}]^\top\) and
\(s = [s^{(B)}, s^{(D)}]^\top\).  The kinetic model operator \(C\) is such that
\(C(\theta) = [C^{(B)}(\theta^{(B)}), C^{(D)}(\theta^{(D)})]^\top\).

The penalty \(R\) used in this work is specifically designed to penalize large
variations in receptor occupancy throughout the brain, exploiting the fact that
drug receptor occupancy can be assumed to be relatively uniform in the brain
\cite{Cunningham2010}.  Other parameters of the joint kinetic model are
regularized using anatomical information obtained from an MRI scan.  The penalty
function is defined as:
\begin{IEEEeqnarray}{rCl}
  R(\theta) &=& \beta_{BP} R_B\left(\BP^{(B)}\right) +
  \beta_{k_2} R_B\left(k_2^{(B)}\right) \label{eq-reg}\\
  && + \beta_{R_1} R_B\left(R_1^{(B)}\right) +
  \beta_{\RO} R_V\left(\RO\right)\IEEEnonumber\\
  && + \beta_{k_2} R_B\left(k_2^{(D)}\right) +
  \beta_{R_1} R_B\left(R_1^{(D)}\right),\IEEEnonumber
\end{IEEEeqnarray}
where \(R_V(\RO)\) is the variance of the receptor occupancy map in the brain and
\(R_B\) is a Bowsher regularizer \cite{Vunckx2012} applied on kinetic parameter
maps \(\BP^{(B)}\), \(k_2^{(B)}\), \(R_1^{(B)}\), \(k_2^{(D)}\) and \(R_1^{(D)}\).  To
reduce the number of parameters to select, \(\beta_{k_2}\) and \(\beta_{R_1}\) are
defined as scaled versions of \(\beta_{BP}\) with fixed scales (based on typical
intensity levels for each kinetic parameter).  Therefore the proposed method
depends on two parameters: \(\beta_{\RO}\), controlling the variance penalty
strength on the \(\RO\) map and \(\beta_{BP}\), controlling the strength of the
Bowsher penalty on the other kinetic parameters.

We propose to use a nested ADMM framework \cite{Boyd2011} to
solve~(\ref{eq-obj-joint-direct}) in order to simplify the optimization
procedure by decomposing it into manageable subproblems.  Accordingly, the
objective function~(\ref{eq-obj-joint-direct}) is transformed into a
constrained optimization problem:
\begin{align}
\label{eq-admm-out}
  \theta^*, x^* = \argmin_{\theta,x}&
    {-\mathcal{L}}(y|x) + R(\theta)\\
    &\text{s.t. } C(\theta) = \Gamma^{-1} x\IEEEnonumber.
\end{align}
Within the ADMM framework, (\ref{eq-admm-out}) is solved by alternating between the
three following updates:
\begin{align}
  x^{(n + 1)}&= \arg\min_{x} -\mathcal{L}(y|x) +
               \frac{\rho}{2} \, L_1(\theta^{(n)}, x, \eta^{(n)})
               \label{eq-admm-x}\\
  \theta^{(n + 1)}&= \arg\min_{\theta} R(\theta) +
                    \frac{\rho}{2} \, L_1(\theta, x^{(n + 1)}, \eta^{(n)})
                    \label{eq-admm-theta}\\
                    \eta^{(n + 1)}&= \eta^{(n)} + C(\theta^{(n + 1)}) -
                    \Gamma^{-1} x^{(n + 1)}
  \label{eq-admm-eta}
\end{align}
where \(L_1(\theta, x, \eta) = \left\|C(\theta) - \Gamma^{-1}x +
\eta\right\|_W^2\) completes the scaled form of the augmented Lagrangian term,
\(\rho\) is a scalar controlling the strength of ADMM constraint term
in~(\ref{eq-admm-out}) and \(\eta\) is an auxiliary ADMM variable.
Equation~\eqref{eq-admm-eta} is a trivial parameter update step.
Equation~\eqref{eq-admm-x} is a penalized image reconstruction problem
with a quadratic penalty, which can be solved using the BSREM algorithm
\cite{Ahn2003}.

Equation~\eqref{eq-admm-theta} describes a penalized kinetic model
fitting, likewise solved using the ADMM algorithm, resulting in the following
constrained problem:
\begin{align}
  \theta^*, \gamma^* = \argmin_{\theta,\gamma}
  & \,R(\theta) +
    \frac{\rho}{2} \, L_1(\gamma, x^{(n + 1)}, \eta^{(n)})\label{eq-admm-in}\\
  &\text{s.t. } h(\theta) = h(\gamma) \IEEEnonumber.
\end{align}
\(h\) is a transform from joint kinetic parameter maps (\(\theta\)) to kinetic
parameters for each scan, i.e. if \(\theta = [\BP^{(B)}, k_2^{(B)}, R_1^{(B)},
\RO, k_2^{(D)}, R_1^{(D)}]^\top\), then \(h(\theta)\) is given by \(h(\theta) =
[\BP^{(B)}, k_2^{(B)}, R_1^{(B)}, \BP^{(B)}\,(1 - \RO), k_2^{(D)},
R_1^{(D)}]^\top\).  The ADMM updates to solve~\eqref{eq-admm-in} are:
\begin{IEEEeqnarray}{rCl}
  \theta^{(p + 1)} &= \arg\min_{\theta}& R(\theta) +
  \frac{\mu}{2} L_2(\theta, \gamma^{(p)}, \nu^{(p)})%
  \label{eq-admm2-theta}\\
  \gamma^{(p + 1)} &= \arg\min_{\gamma}&
  \frac{\rho}{2} \, L_1(\gamma, x^{(n + 1)}, \eta^{(n)}) \IEEEnonumber\\
  && + \frac{\mu}{2} L_2(\theta^{(p + 1)}, \gamma, \nu^{(p)})%
  \label{eq-admm2-gamma}\\
  \nu^{(p + 1)} &= \nu^{(p)} + &\theta^{(p + 1)} - \gamma^{(p + 1)},
\end{IEEEeqnarray}
where \(L_2(\theta, \gamma, \nu) = \left\| h(\theta) - h(\gamma) +
\nu\right\|_{\Lambda}^2\) completes the scaled form of the augmented Lagrangian,
\(\mu\) is the ADMM relaxation parameter and \(\nu\) is an ADMM auxiliary variable.
\(\Lambda\) is a weight used to equalize the contribution of the different kinetic
parameters in the augmented Lagrangian term (based on typical value of each
kinetic parameter).  Equation~\eqref{eq-admm2-theta} resembles an image
denoising problem and is solved using the nonlinear conjugate gradient algorithm
with a Newton-Raphson line search.  Finally,
equation~\eqref{eq-admm2-gamma} describes a penalized kinetic fitting
problem that can be parallelized over image pixels.  It is solved using a
modified version of the basis function method described in \cite{Gunn1997},
further described in Supplemental Material~\ref{tmi_fit_reg_solver}.

\subsection{Implementation details}
\label{tmi_method_implementation}
This section describes techniques used to improve the performance of the
proposed algorithm.  In order to reduce the computational requirements, the
proposed method was applied slice by slice using the single slice rebinning
(SSRB) algorithm \cite{Defrise1997}.  Extension to fully three-dimensional
reconstruction is straight-forward.  The geometrical component of the PET system
matrix was implemented using Siddon's ray tracing \cite{Siddon1985a,Jacobs1998}.
Due to the nonlinear nature of the forward model, initialization of the
reconstruction algorithm is critical.  In this work, the ADMM algorithm was
initialized by the result of conventional indirect estimation: OSEM followed by
a 4 mm Gaussian filter \cite{Germino2018} and SRTM fitting.

Parameter selection is critical to the performance of Bayesian reconstruction
algorithms.  The proposed method requires the selection of a few parameters,
namely the regularization strengths (\(\beta_{BP}\), \(\beta_{RO}\), etc. in
\eqref{eq-reg}) as well as ADMM parameters \(\rho\) and \(\mu\), which control the
weight of the augmented Lagrangian penalty term \cite{Boyd2011}.  Here, they
were selected by balancing the cost terms in the ADMM subproblems and by
inspecting the convergence behavior of each ADMM subproblem (this was performed
on a single noise realization later excluded from the phantom evaluation).  As
described in Section~\ref{tmi_method_admm}, the \(\beta\) regularization
parameters were grouped such that the only parameters to select were
\(\beta_{BP}\) and \(\beta_{RO}\).  The resulting set of parameters used for images shown in Fig.~\ref{fig-bv-real-bp_ro} is summarized in Supplemental Material Table~\ref{tbl-jdpr-params}.

In practice, direct reconstruction methods can be degraded by background
structures (e.g. skull, neck) present in the field of view that may not adhere
to the underlying kinetic model as well as by outliers in PET reconstructed
images and kinetic modeling.  In this work, the background activity issue for
\emph{in-vivo} data was addressed by only estimating kinetic parameters in the brain
region.  Dynamic activity values from the skull and other background structures
were estimated from an initial OSEM reconstruction and added to the synthesized
brain TACs at each iteration.  Outliers in indirect estimation methods were
handled by clipping kinetic parameter maps to a fixed range.  While infrequent
for direct methods, pixel updates outside of the same range were omitted to
prevent extreme values from appearing in intermediate dynamic PET images.

\subsection{Numerical simulation phantom}
\label{tmi_data_sim}
A numerical phantom was used to quantitatively evaluate the proposed method.
The numerical phantom was constructed from the non-human primate (NHP) NIH
macaque template~(NMT) \cite{Jung2021,Seidlitz2018}, which was used to
generate a brain atlas for PET comprised of 13 regions.  Kinetic parameters were
assigned to each region of the phantom based on SRTM analysis of \emph{in-vivo} NHP
[\textsuperscript{11}C]MK-6884 PET scans.  Noiseless dynamic PET images were
then synthesized with the SRTM model using the gray matter cerebellum as
reference region \cite{Li2022}.  To mimic an occupancy study, changes in \(\BP\)
were introduced between the baseline and post-drug scan, corresponding to a
receptor occupancy \(\RO\) following the Hill model in (\ref{eq-hill}).  Two simulations
were used: one where the \(\RO\) map was chosen to be uniform across brain regions
and one where the \(\RO\) map varies in a brain region.  Brain regions of the
atlas as well as representative time activity curves are shown in
Fig.~\ref{fig-phant}.

Dynamic PET projections were then obtained by forward-projecting the noise-free
dynamic images mimicking the geometry of a Discovery MI PET/CT scanner (GE
Healthcare).  Poisson deviates were applied to the obtained PET sinograms to
approach the noise level in a typical \emph{in-vivo} NHP scan.

The proposed method was evaluated via bias-variance analysis.  The proposed
joint direct parametric reconstruction (`JDPR') was compared to standard
indirect methods: indirect parametric reconstruction (`IPR') consisting in OSEM
reconstruction of dynamic PET images followed by SRTM fitting using basis
functions \cite{Gunn1997} and OSEM reconstruction of dynamic PET images with
post-reconstruction Gaussian smoothing followed by SRTM fitting (`IPRs').
Additionally the proposed method was compared to a method representative of
existing direct parametric reconstruction methods (`DPR'), i.e., replacing the
joint penalty on receptor occupancy (\(\beta_{RO} = 0\) in \eqref{eq-reg}) by
Bowsher regularization of post-drug binding potential.  DPR was applied to each
scan separately, before estimating RO.  For this analysis, pixel-wise bias and
standard deviation were calculated using:
\begin{align}
  \mathrm{bias} &= \frac{1}{K} \sum_{k=1}^K \theta^{(k)} - \hat{\theta}\\
  \mathrm{std} &= \frac{1}{K - 1} \sum_{k=1}^K (\theta^{(k)} - \bar{\theta})^2,
\end{align}
where \(K\) is the number of noise realizations, \(\theta^{(k)}\) is a set of
reconstructed kinetic parameter maps for noise realization \(k\), \(\hat{\theta}\)
is the set of ground truth kinetic parameter maps and \(\bar{\theta} =
\frac{1}{K} \sum_{k=1}^K \theta^{(k)}\).  In order to avoid extreme,
physiologically implausible values for estimated kinetic parameters that may
severely impact bias and variance estimates, reconstructed kinetic parameter
maps were first clipped.  Bias-variance curves were obtained for each method by
varying reconstruction parameters: for IPR, the reconstruction parameter was the
number of iterations; for IPRs, the Gaussian filter full width at half maximum
(FWHM); for DPR, the regularizer strength \(\beta_{BP}\) was varied to obtain the
curve; finally, for the proposed method (JDPR), the regularizer strength
\(\beta_{BP}\) and \(\beta_{RO}\) were used to obtain the bias-variance curve.

\subsection{In-vivo studies}
\label{tmi_data_invivo}
The proposed method was applied to data acquired in two non-human primates (with
weights ranging from 12.2 kg to 13.4 kg and from 15.1 kg to 17.7 kg across
studies) using [\textsuperscript{11}C]MK-6884, a muscarinic acetylcholine
receptor 4 (M4) ligand and CVL-231, an M4 positive allosteric modulator under
development.  All acquisitions were performed on a GE PET/CT Discovery MI
scanner. The study protocol, described in detail in \cite{Belov2024}, included
two dynamic PET scans: one at baseline and the other under a blocking condition
with intravenous administration of CVL-231 by bolus-plus-infusion (loading dose
of 0.8 mg/kg given 10 min before radiotracer followed by constant infusion of
varying concentration--0.25, 0.5, 1, 1.7 and 3.4 mg/kg--for a 90 minute
administration protocol).  Dynamic studies were truncated to 45 minutes.  In
total, 7 pairs of scans were performed at different CVL-231 concentrations.
Baseline and post-drug studies were performed with the same field of view and
subject position, therefore registration between scans was not required.  In the
absence of ground truth, the proposed method was compared in terms of agreement
with baseline methods in the striatum, a region with a higher M4 receptor
density and therefore higher signal-to-noise ratio.

\section{Results}
\label{tmi_results}
\begin{figure}[htbp]
\centering
\includegraphics[trim=0.1in 0.2in 0.1in 0.1in]{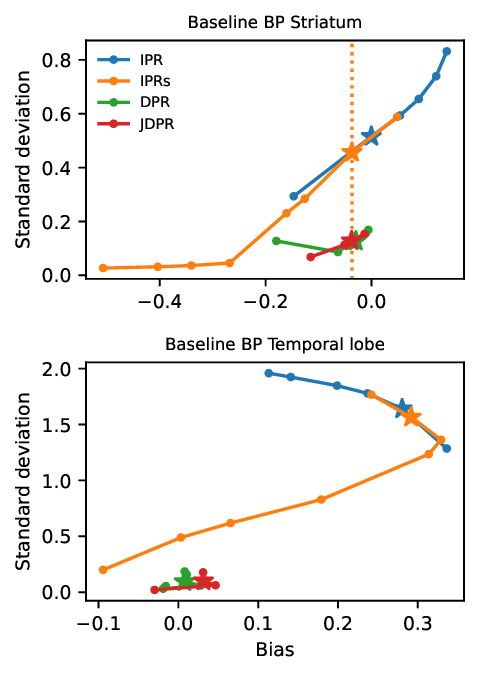}
\caption{\label{fig-bv-bp}Bias/standard deviation plots for baseline \(\BP\) estimation averaged in the striatum (top) and temporal lobe (bottom) regions.  Each plot was obtained by varying a reconstruction parameter for each method (varying the iteration number for IPR, the post-reconstruction Gaussian filter width for IPRs, $\beta_{BP}$ for DPR and JDPR).  The vertical orange dashed line indicates the reference reconstruction setting for the IPRs method (1.5 mm FWHM filter) used for the bias/standard deviation maps shown in Fig.~\ref{fig-bv-map-bp}.  The star markers for each method indicate the selected parameter best approaching the reference reconstruction setting in the striatum.}
\end{figure}

\begin{figure}[!h]
\centering
\includegraphics[trim=0.1in 0.2in 0.1in 0.1in]{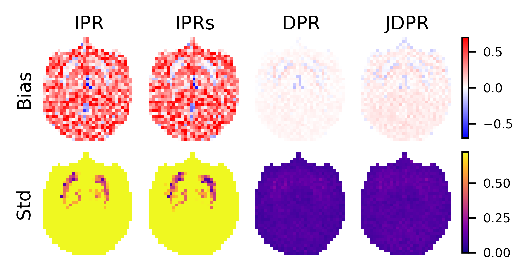}
\caption{\label{fig-bv-map-bp}Bias (top) and standard deviation (bottom) maps for \(\BP\) estimation for the different methods.  Reconstruction settings for each method are indicated by stars in Fig.~\ref{fig-bv-bp}.}
\end{figure}

\begin{figure*}[!htb]
\centering
\includegraphics[trim=0.1in 0.2in 0.1in 0.1in]{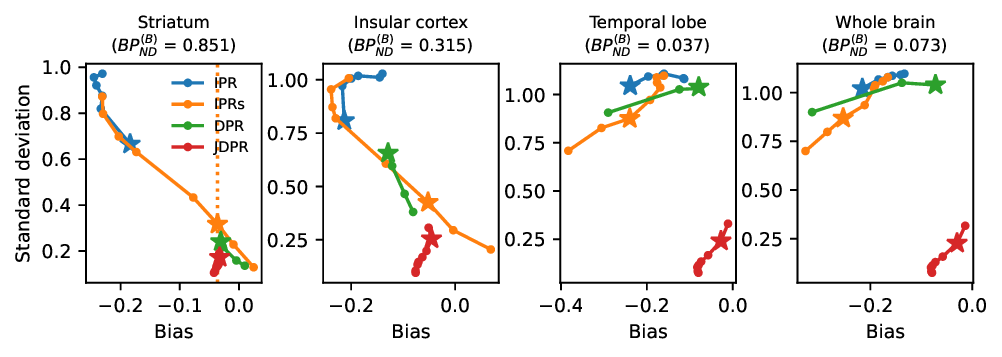}
\caption{\label{fig-bv-ro}Bias/standard deviation plots for RO estimation averaged in, from left to right, the striatum, insular cortex, temporal lobe and whole brain. The true parameter value is indicated in the title.  Each plot was obtained by varying a reconstruction parameter for each method (varying the iteration number for IPR, the post-reconstruction Gaussian filter width for IPRs, $\beta_{BP}$ for DPR and $\beta_{RO}$ for JDPR).  The vertical orange dashed line indicates the reference reconstruction setting for the IPRs method (4 mm FWHM filter).  The star markers, for each method the selected parameter best approaching the reference reconstruction setting in the striatum.}
\end{figure*}

\begin{figure}[htbp]
\centering
\includegraphics[trim=0.1in 0.2in 0.1in 0.1in]{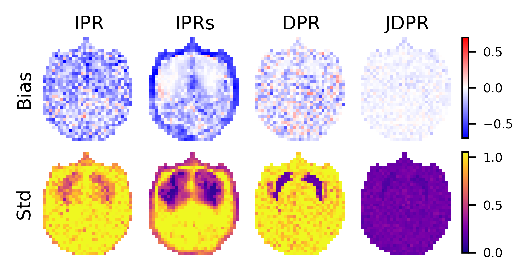}
\caption{\label{fig-bv-map-ro}Bias (top) and standard deviation (bottom) maps for RO estimation for different reconstruction methods with settings indicated by stars in Fig.~\ref{fig-bv-ro}.}
\end{figure}

\begin{figure}[htbp]
\centering
\includegraphics[trim=0.1in 0.2in 0.1in 0.1in]{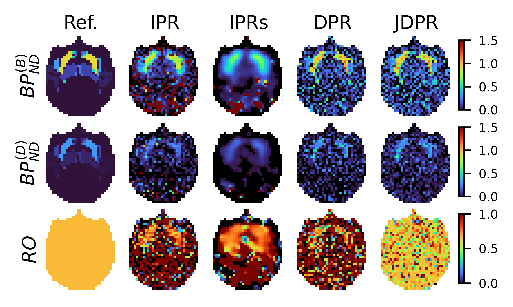}
\caption{\label{fig-bv-real-bp_ro}Baseline binding potential (top) and receptor occupancy maps for a single noise realization and settings indicated by stars in Fig.~\ref{fig-bv-ro}.  Values are clipped to the intensity range displayed in the color bar (pixels with out-of-range low values are in black, high values are in dark red).}
\end{figure}

\begin{figure}[!htpb]
\centering
\includegraphics[trim=0.1in 0.15in 0.1in 0.1in]{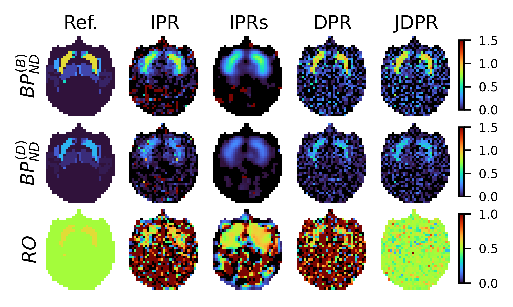}
\caption{\label{fig-bv-ro_var_real}Binding potential (top) and receptor occupancy maps for a single noise realization with spatially varying receptor occupancy, and settings indicated by stars in Fig.~\ref{fig-bv-ro_var}.  Values are clipped to the intensity range displayed in the color bar (pixels with out-of-range low values are in black, high values are in dark red).}
\end{figure}

\begin{figure}[!h]
\centering
\includegraphics[trim=0.1in 0.15in 0.1in 0.255in]{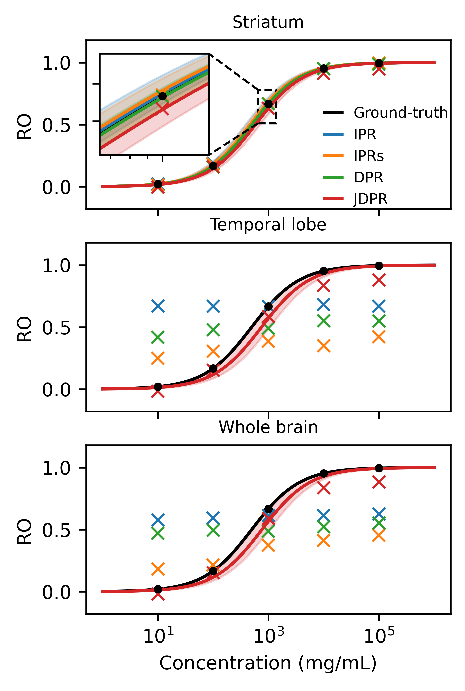}
\caption{\label{fig-sim-hill}Estimated receptor occupancy for each method (crosses) as a function of simulated drug concentration compared to ground truth Hill curve used for the simulation.  The shaded region shows the 20\%-80\% percentiles of the fitted Hill curve over 50 noise realizations, in each region, excluding outliers.  The proposed method best approaches the ground-truth curve, especially in low binding regions. Note that the envelope region is omitted for IPR, IPRs and DPR outside of the striatum, where these methods yield unreliable RO estimates.}
\end{figure}

\subsection{Simulation phantom}
\label{sec:orgb3e79fe}

The phantom described in Section~\ref{tmi_data_sim} was used to compare the
different reconstruction methods.  For this evaluation, the receptor occupancy
was fixed to 66\% (following~(\ref{eq-hill}) with \(E_{max} = 1\),
\(\frac{ED_{50}}{c} = \frac{1}{2}\)) in the whole brain and 50 noise realizations
were reconstructed.  The proposed joint direct method was compared to
conventional parametric estimation: (1) IPR with 12 OSEM subsets, varying the
number of iterations, (2) IPRs with a fixed number of iterations and subsets
varying Gaussian filter FWHMs and (3) DPR varying \(\beta_{\BP}\).  Both the
direct and proposed joint-direct methods used a Huber spatial prior
\cite{Yu2002}.  Methods were compared in the striatum (a high-binding region for
this particular tracer), the temporal lobe (a low binding region) and the whole
brain.

\begin{figure*}[!h]
\centering
\includegraphics[trim=0.1in 0.2in 0.1in 0.1in]{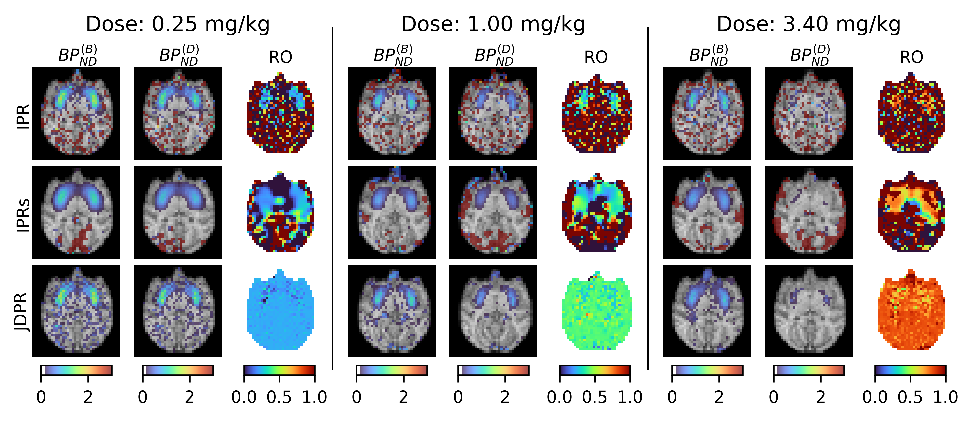}
\caption{\label{fig-hill_invivo_kin}Kinetic parameter images for the NHP [\textsuperscript{11}C]MK-6884 occupancy study at different drug concentrations.  For each concentration, the baseline, post-drug BP (overlaid on the MR image), and the RO map are shown for each method: (top) IPR, (middle row) IPRs with 4 mm Gaussian filter and (bottom) proposed JDPR method.  For all concentrations, the proposed method results in improved RO estimates, especially in low binding regions.}
\end{figure*}

The bias/standard deviation plots for baseline \(\BP\) estimation are shown in
Fig.~\ref{fig-bv-bp}.  The curves were obtained by varying reconstruction parameters controlling the bias/standard deviation trade-off, i.e., varying the iteration number for IPR (from 1 to 9 iterations), the post-reconstruction Gaussian filter width for IPRs (from 1 to 7 mm), $\beta_{BP}$ for DPR and JDPR (from $\beta_{BP} = 1$ to $\beta_{BP} = 500$).
The curves demonstrate the improvement of both DPR and proposed JDPR over
indirect estimation methods.  As expected, in the striatum as well as in low
binding regions, the proposed and direct methods both achieved consistently low
bias, as well as lower standard deviation at a given bias level than the
conventional methods.  The proposed method, as well as direct reconstruction,
were able to match this low bias while reducing standard deviation three-fold
compared to the reference reconstruction indicated by the vertical orange dotted
line.  In this study, the reference reconstruction setting was selected
arbitrarily to result in low bias in the striatum, using a Gaussian filter with
FWHM 1.5 mm.  Bias and variance maps shown on Fig.~\ref{fig-bv-map-bp}
illustrate the dramatic improvement both in bias and standard deviation as
compared to conventional indirect methods.

Fig.~\ref{fig-bv-ro} presents bias and standard deviation plots of \(\RO\)
estimates when varying reconstruction parameters for each method.
The curves were obtained by varying reconstruction parameters controlling the bias/standard deviation trade-off, i.e., varying the iteration number for IPR, the post-reconstruction Gaussian filter width for IPRs, $\beta_{BP}$ (from $\beta_{BP} = 1$ to $\beta_{BP} = 100$) for DPR and $\beta_{\RO}$ (from $\beta_{\RO} = 500$ to $\beta_{\RO} = 10000$) for JDPR.
In the striatum, the proposed method approached the performance of the
conventional approach used to estimate receptor occupancy, i.e. OSEM
reconstruction followed by Gaussian filtering (FWHM = 4 mm) and SRTM fitting
(indicated by the vertical orange dashed line in the figure).  The proposed
method offered a reduction in standard deviation for equal bias for \(\RO\)
estimates compared to the conventional methods (40\% reduction over IPRs for the
operating points marked by a star).  Direct reconstruction likewise achieved a
low bias and low standard deviation in the striatum.  The improvement afforded
by JDPR was much more pronounced in brain regions with intermediate to lower receptor densities, where IPR, IPRs and DPR methods introduced a
large bias (over 20\% for indirect methods) with high variance, whereas the
proposed method resulted in low bias and variance, consistent with results
obtained in the striatum.  Corresponding bias and variance maps for the
reconstruction settings marked by a star in Fig.~\ref{fig-bv-ro} are shown in
Fig.~\ref{fig-bv-map-ro}, while Fig.~\ref{fig-bv-real-bp_ro} presents
representative examples of \(\BP\) and RO images for a single noise realization.
They confirm the improvement offered by the proposed method, which matched the
low bias observed in the reference reconstruction (IPRs) while dramatically
reducing the bias in other regions.  A similar reduction was observed on the
standard deviation map, confirming that the proposed method can yield more
accurate and robust estimates.  Figure~\ref{fig-bv-real-bp_ro} also shows improved $\BP$ estimation for the DPR and JDPR methods. This is a result of the direct estimation approach, which improves the precision of $\BP$ estimates.  An example of ADMM cost evolution over iterations is shown in
Supplemental Material Fig.~\ref{fig-admm-cost}.

The proposed method penalizes variance in the receptor occupancy map, based on
the commonly used assumption that blocking drug effects are mostly uniform
across the brain.  In order to evaluate the performance of the proposed
algorithm when this assumption is not valid, a numerical simulation was carried
out with a nonuniform receptor occupancy map, set at 60\% in striatum and 50\% in
the rest of the brain.

Representative results for a single noise realization shown in
Fig.~\ref{fig-bv-ro_var_real} demonstrate the ability of JDPR to capture
spatial variations in receptor occupancy.  Bias-variance analysis reported in
Supplemental Material (Figs.~\ref{fig-bv-ro_var} and~\ref{fig-bv-ro_var_map})
shows that the proposed method achieved a similar level of bias and variance as
for the uniform RO case.

Finally, numerical simulations were performed for different drug
concentration--and thus occupancy--levels.  The simulated receptor occupancy
followed the Hill equation in (\ref{eq-hill}).  Fifty noise realizations were
simulated for each concentration level.  RO values were averaged in each region
for each method, and for each noise realization, the Hill equation was fitted to
the regional RO estimates plotted as function of the simulated concentration
level.  The receptor occupancy values estimated by each method are shown, along
with the Hill curve fits, in Fig.~\ref{fig-sim-hill}.  The shaded curve
envelopes correspond to the 20\%-80\% percentiles across noise realizations.
A summary of the RO estimation performance, measured in term of bias and standard deviation is reported in Supplemental Material Table~\ref{tbl-hill-sim-bv}.  In the striatum, all methods resulted in good
RO estimation, the proposed method yielding the lowest variance, at the expense of a slight loss of accuracy (around 5\%).  In other regions,
typically with lower binding and thus SNR, conventional methods failed to
capture the shape of the Hill curve owing to large errors in RO estimation
(the confidence intervals in Fig.~\ref{fig-sim-hill} were omitted for the IPR, IPRs and DPR in non-striatal regions since the fitted RO values are severely biased, preventing direct comparison with JDPR).  By contrast,
the proposed method consistently estimated RO for all concentrations with high
accuracy and low variance, with a slight error increase for high receptor
occupancy in low binding regions.  This is confirmed by the high R-squared values ($>$ 90\%) reported in Supplemental Material Table~\ref{tbl-hill-sim-bv}.

\subsection{\emph{In-vivo} experiments}
\label{sec:orgc0d827a}

\begin{figure}[htbp]
\centering
\includegraphics[trim=0.1in 0.15in 0.1in 0]{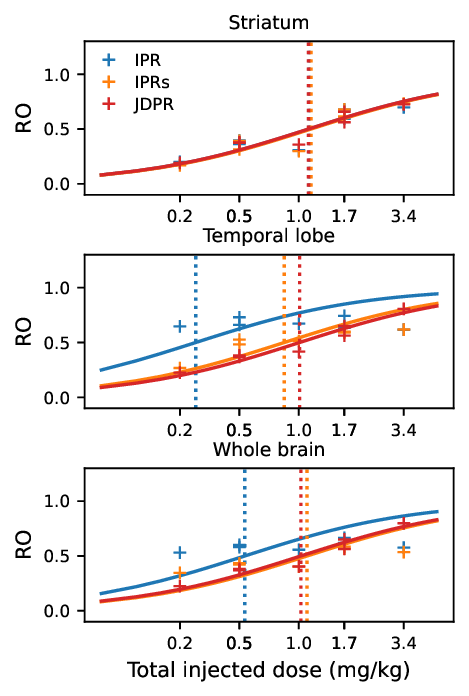}
\caption{\label{fig-hill_invivo}Estimated RO values and fitted Hill curve for \emph{in-vivo} data in different brain regions.  Dashed vertical lines indicate the estimated \(ED_{50}\).  While all methods result in estimated RO values in agreement with the logistic Hill curve in the striatum, only the proposed method achieves a similar performance in other brain regions.}
\end{figure}

The proposed method was evaluated on the \emph{in-vivo} data described in
Section~\ref{tmi_data_invivo}.  Representative kinetic parameter maps are
shown in Fig.~\ref{fig-hill_invivo_kin}.  In this study, \(\beta_{BP}\) was
selected such that the proposed joint direct reconstruction matches the bias
level of the \(\BP\) parameter achieved by the conventional indirect method in the
striatum.  \(\beta_{RO}\) was selected to match the RO bias of the conventional
indirect method in the striatum.  The proposed method resulted in improved RO
estimation, especially in low binding regions.  The estimation of baseline \(\BP\)
was also improved, achieving a contrast recovery in the striatum similar to the
IPR method with a reduced noise level.  The reduction in variance also
translated into improved fitting of the Hill curve (\ref{eq-hill})
(Fig.~\ref{fig-hill_invivo}).  While no ground truth RO was available for
\emph{in-vivo} data, the proposed method resulted in RO estimates matching with the
logistic curve, suggesting that the estimated values are consistent with the
model.  Additionally, the \(ED_{50}\) estimate for JDPR in all brain regions was
consistent with that obtained in the striatum, enabling estimation of RO from
the whole brain, which, we hypothesize, will give more power to receptor
occupancy studies.

\section{Discussions}
\label{tmi_discussions}
The proposed joint-direct reconstruction framework relies on the minimization of
an objective function linking PET projection measurements to kinetic parameter
maps and including a joint prior on the baseline and post-drug scans.  This
formulation achieves two objectives: (1) bias control in estimation of kinetic
parameters thanks to the log-likelihood term and (2) application of a variance
penalty on receptor occupancy distribution (resulting from the assumption of
uniform drug exposure in the brain).  This enables accurate and precise
estimation of RO in regions with lower receptor density from all the brain
regions.  From Fig.~\ref{fig-bv-ro}, it is worth noting that, in the striatum,
both IPRs and DPR could achieve ideal RO estimation with near zero bias and
standard deviation.  This is a consequence of the RO calculation where biases in
\(\BP\) estimates from baseline and post-drug scans cancel each other (which
explains the shape of the curve where reduction in variance is accompanied by
reduction in bias, contrary to the more usual trend seen, e.g. in
Fig.~\ref{fig-bv-bp}).  In this work, the parameter for IPRs used as reference was selected to achieve a reasonable balance between RO and BP bias.

Since the variance penalty is implemented as a soft constraint, the proposed
method can relax the hypothesis of uniform receptor occupancy \cite{deLaat2021}
as shown in Fig.~\ref{fig-bv-ro_var_real} and Supplemental Material
Figs.~\ref{fig-bv-ro_var} and~\ref{fig-bv-ro_var_map}.
Therefore, the proposed method is able to exploit RO uniformity when relevant but still provide the benefits of direct parametric mapping and joint RO estimation when the uniform RO constraint is relaxed.  This could
happen, for instance, in cases where the radiotracer binds to more targets than
the drug under investigation.

While the reported evaluations performed on simulations and in-vivo data did not
include motion in the measurements, incorporation of rigid motion is
straightforward in the proposed estimation framework.  It only requires the
addition of a rigid deformation operator in the forward
model~(\ref{eq-fwd-rec}), resulting in increased computational cost, but
otherwise no change to the estimation framework, which, we expect, will enable
scanning over multiple days, without need for sedation.

One limitation of the evaluation is the tracer used for the study.
[\textsuperscript{11}C]MK-6884 is characterized by high binding in the striatum,
low binding in most other regions of the brain and few regions with intermediate
level of binding.  In future work, the proposed method will be applied to other
tracers and drug candidates to evaluate its performance with different tracer
distributions.  The simulation results give an indication of the performance of proposed method in regions with intermediate levels of binding. As shown in Fig.~\ref{fig-bv-ro}, the insular cortex region was assigned an intermediate $\BP$ value ($BP=0.315$), and bias/standard deviation curves show the improvement offered by JDPR over conventional methods. The proposed method may offer diminishing benefits for tracers with high binding in the whole brain. In this case, the method may still be beneficial for low count data.

Another limitation is the number of hyperparamaters to select, namely the regularization strengths and ADMM auxiliary variables. In practice, the regularization parameters are coupled based on the expected dynamic ranges as described in Section~\ref{tmi_method_implementation} and selected by targetting a fixed ratio between log-likelihood and regularization. ADMM parameters ($\rho$ and $\mu$) are selected from a single iteration to minimize the cost of the corresponding subproblems.

\section{Conclusion}
\label{sec:org6cfeedd}

This paper proposes a joint direct reconstruction for receptor occupancy
estimation from baseline and post-drug PET scans.  The proposed method combines
direct parametric estimation with a prior on the receptor occupancy map
estimated from both scans.  Validation on numerical simulations demonstrated
improved the bias variance trade-off compared to traditional indirect as well as
non-joint direct methods, especially in regions with low receptor densities
which are often excluded from occupancy analyses due to the poor estimation
performance.  The proposed method was successfully applied to \emph{in-vivo} MK-6884
data to characterize the relationship between drug concentration and receptor
occupancy.  The reduction in estimation variance achieved with the proposed
method could be used to reduce the sample size required for occupancy studies
and/or decrease radiation exposure in study subjects for tracers with low and intermediate binding regions.

\FloatBarrier

\label{sec:org45b9209}

\bibliographystyle{IEEEtran_tmi}
\bibliography{IEEEabrv,bstoptions,bibliography}

\clearpage
\onecolumn

\section*{Supplemental material}
\label{sec:orgb79b4cb}
\renewcommand{\thepage}{S.\arabic{subsection}}
\renewcommand{\thesubsectiondis}{S.\arabic{subsection}}
\renewcommand{\thesubsection}{S.\arabic{subsection}}
\setcounter{equation}{0}
\setcounter{page}{1}
\renewcommand{\thefigure}{S.\arabic{subsection}.\arabic{figure}}
\renewcommand{\theequation}{S.\arabic{subsection}.\arabic{equation}}

\subsection{Penalized kinetic fitting using basis functions}
\label{tmi_fit_reg_solver}
\setcounter{figure}{0}
\setcounter{table}{0}

This section describes the method used to
minimize~\eqref{eq-admm2-gamma}.  The problem is reformulated into the
following generic form:
\begin{align}
\label{eq-fit-pen-gamma}
  \gamma^* = \arg\min_{\gamma}
  \frac{\rho}{2} \left\|z - C(\gamma)\right\|_W^2 +
  \frac{\mu}{2} \left\|\zeta - h(\gamma)\right\|_{\Lambda}^2,
\end{align}
where \(z\) is the concatenation of the dynamic PET images for baseline and
post-drug scans and \(\zeta\) is the set of kinetic parameters:
\(\zeta=[\zeta_{\BP^{(B)}}, \zeta_{k_2^{(B)}}, \zeta_{R_1^{(B)}},
\zeta_{\BP^{(D)}}, \zeta_{k_2^{(D)}}, \zeta_{R_1^{(D)}}]\).

The problem in (\ref{eq-fit-pen-gamma}) is modified to enable the use of the basis
function technique described in \cite{Gunn1997}.  The strategy consists in
extracting the nonlinear components of the operators in quadratic terms into
basis functions and express the overall optimization problem as a sum of
quadratic terms with linear operators for each basis function.  The kinetic
operator \(C(\gamma)\) can be reformulated as in \cite{Gunn1997}, taking into a
account the two scans, resulting in:
\begin{IEEEeqnarray}{rCCC}
  C(\gamma) &=& \left[\begin{matrix}
      E_{\kappa^{(B)}} & F_{\kappa^{(B)}} & 0 & 0\\
      0 & 0 & E_{\kappa^{(D)}} & F_{\kappa^{(D)}}
    \end{matrix}\right] \,&
  \left[\begin{matrix}
      k_2^{(B)} \\ R_1^{(B)} \\ k_2^{(D)} \\ R_1^{(D)}
    \end{matrix}\right]\IEEEnonumber\\
  &\triangleq& \mathcal{C}_{\kappa} \,& \tilde{\gamma},
\end{IEEEeqnarray}
where \(\tilde{\gamma} = [k_2^{(B)}, R_1^{(B)}, k_2^{(D)}, R_1^{(D)}]\) and the
basis functions are expressed as \(E_{\kappa} = C_R(t) \ast \exp\left(-\kappa
t\right)\) and \(F_{\kappa} = C_R - \kappa \, E_{\kappa}\) with
\(\kappa=\frac{k_2}{1 + \BP}\).  Basis functions are precalculated for pairs of
\(\kappa = (\kappa^{(B)}, \kappa^{(D)})\) values.

The second quadratic term can be reformulated as:
\begin{IEEEeqnarray}{rCCCCl}
  \zeta - h(\gamma) &=&\left[\begin{matrix}
      \zeta_{\BP^{(B)}} + 1\\
      \zeta_{k_2^{(B)}}\\
      \zeta_{R_1^{(B)}}\\
      \zeta_{\BP^{(D)}} + 1\\
      \zeta_{k_2^{(D)}}\\
      \zeta_{R_1^{(D)}}
    \end{matrix}\right] &-&
  \left[\begin{matrix}
      \frac{1}{\kappa^{(B)}} & 0 & 0 & 0\\
      1 & 0 & 0 & 0\\
      0 & 1 & 0 & 0\\
      0 & 0 & \frac{1}{\kappa^{(D)}} & 0\\
      0 & 0 & 1 & 0\\
      0 & 0 & 0 & 1
    \end{matrix}\right] \,&
  \left[\begin{matrix}
      k_2^{(B)} \\ R_1^{(B)} \\ k_2^{(D)} \\ R_1^{(D)}
    \end{matrix}\right]\IEEEnonumber\\
  &\triangleq& \tilde{\zeta} &-& H_{\kappa} \,& \tilde{\gamma},
\end{IEEEeqnarray}
where we used \(\BP = \frac{k_2}{\kappa} - 1\).

The transformed objective function is:
\begin{IEEEeqnarray}{rCl}
  \tilde{\gamma}_{\kappa}^* = \arg\min_{\gamma}
  \frac{\rho}{2} \left\|z - \mathcal{C}_{\kappa}\,\tilde{\gamma}\right\|_W^2 +
  \frac{\mu}{2} \left\|\tilde{\zeta} - H_{\kappa}\,\tilde{\gamma}
  \right\|_{\Lambda}^2,
\label{eq-obj-kin-pen-linear}
\end{IEEEeqnarray}
which can now be solved analytically:
\begin{IEEEeqnarray}{rCl}
  \tilde{\gamma}_{\kappa}^* = \left(
    \rho C_{\kappa}^\top W C_{\kappa} +
    \mu H_{\kappa}^\top \Lambda H_{\kappa}\right)^{-1}
  \left(
    \rho C_{\kappa}^\top W z +
    \mu H_{\kappa}^\top \Lambda \tilde{\zeta} \right).
\end{IEEEeqnarray}
After calculating \(\tilde{\gamma}_{\kappa}^*\) for a grid of \((\kappa^{(B)},
\kappa^{(D)})\) pairs, the final step of the optimization procedure consists in
finding the pair resulting in the smallest cost in (\ref{eq-obj-kin-pen-linear}) and
use the corresponding \(\tilde{\gamma}_{\kappa}^*\) to assign the \(k_2\) and \(R_1\)
kinetic parameters, thus solving~(\ref{eq-fit-pen-gamma}).

\clearpage
\newpage

\subsection{ADMM convergence}
\label{sec:org2ebaaf1}
\setcounter{figure}{0}
\setcounter{table}{0}
\renewcommand{\thefigure}{S.\arabic{subsection}}

\begin{figure}[htbp]
\centering
\includegraphics{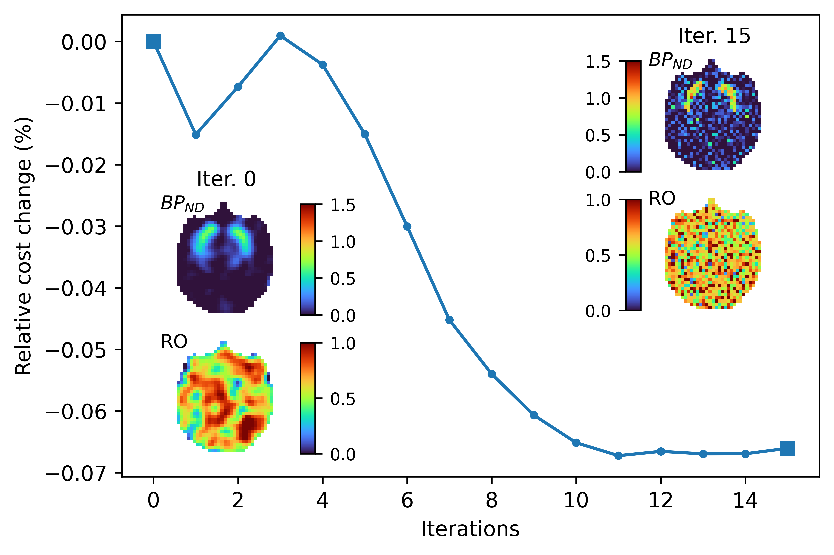}
\caption{\label{fig-admm-cost}ADMM cost as function of iterations for single noise realization with parameters used in Fig.~\ref{fig-bv-real-bp_ro}.  Parametric maps (\(\BP\) and RO) are shown for the initial estimate and final iteration.}
\end{figure}

\clearpage
\newpage

\subsection{Simulation with varying receptor occupancy}
\label{sec:org65cdb9d}
\setcounter{figure}{0}
\setcounter{table}{0}
\renewcommand{\thefigure}{S.\arabic{subsection}.\arabic{figure}}

\begin{figure}[!htpb]
\centering
\includegraphics[trim=0.1in 0.15in 0.1in 0.1in]{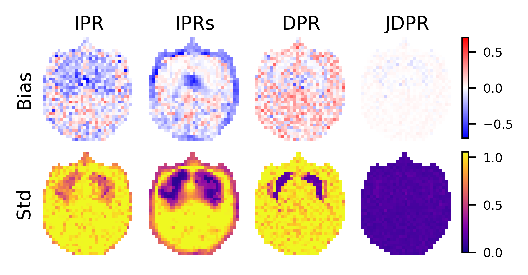}
\caption{\label{fig-bv-ro_var_map}Bias/standard deviation maps for images indicated by stars in Fig~\ref{fig-bv-ro_var}.}
\end{figure}

\begin{figure*}[!htpb]
\centering
\includegraphics[trim=0.1in 0.15in 0.1in 0.1in]{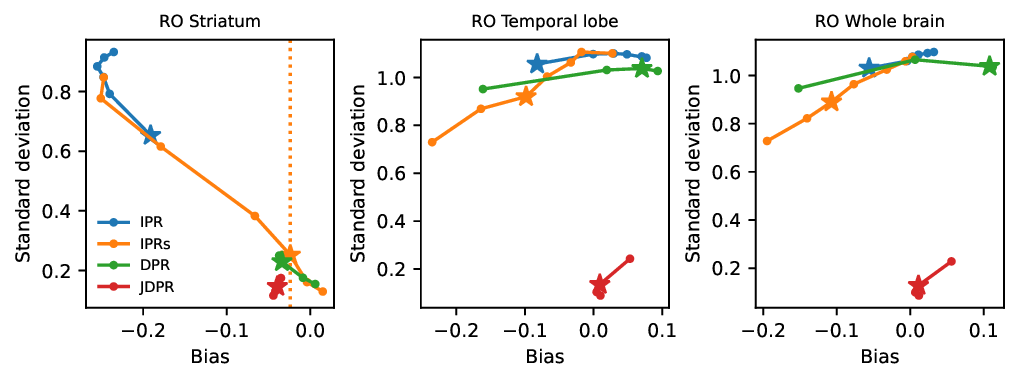}
\caption{\label{fig-bv-ro_var}Bias/standard deviation plots for RO estimation averaged in the striatum (left), temporal lobe (middle) and whole brain (right) regions for the base of a nonuniform RO distribution (RO = 0.6 in the striatum and 0.5 in the rest of the brain).  The vertical orange dashed line indicates the reference reconstruction setting for the IPRs method (4 mm FWHM filter).  The star markers, for each method the selected parameter best approaching the reference reconstruction setting.}
\end{figure*}

This section reports the results of the numerical simulation performed assuming
a spatially varying receptor occupancy.  Figure~\ref{fig-bv-ro_var_map} shows
bias and standard deviation maps for different methods and
Fig.~\ref{fig-bv-ro_var} shows the bias/standard deviation plots for varying
reconstruction parameters.

\clearpage
\newpage

\subsection{Reconstruction parameters}
\label{sec:org0f9fbdb}
\setcounter{figure}{0}
\setcounter{table}{0}
\renewcommand{\thetable}{S.\arabic{subsection}}

\begin{table}[htbp]
\caption{\label{tbl-jdpr-params}Reconstruction parameters for JDPR method: for simulations, parameters used to obtain results indicated by a star in Fig.~\ref{fig-bv-ro}; for in-vivo reconstruction: parameters used to obtain results in Fig.~\ref{fig-hill_invivo_kin} (1 mg/kg).}
\centering
\begin{tabular}{lrr}
Parameter & Simulation & In-vivo\\[0pt]
\hline
\(\beta_0\) & 10 & 0.01\\[0pt]
\(\beta_{RO}\) & 1000 & 1\\[0pt]
\(\rho\) & 5 & 1\\[0pt]
\(\mu\) & 100 & 1\\[0pt]
Outer ADMM iterations (Eqs. 13, 14, 15) & 15 & 8\\[0pt]
Inner ADMM iterations (Eqs. 17, 18, 19) & 6 & 6\\[0pt]
CG iterations (Eq. 17) & 4 & 4\\[0pt]
BSREM iterations & 4 & 4\\[0pt]
BSREM subsets & 12 & 12\\[0pt]
\end{tabular}
\end{table}

\clearpage
\newpage

\subsection{Bias variance of RO estimates over concentration (simulation)}
\label{sec:orgecd1a58}
\setcounter{figure}{0}
\setcounter{table}{0}
\renewcommand{\thetable}{S.\arabic{subsection}}

\begin{table}[htbp]
\caption{\label{tbl-hill-sim-bv}Bias/standard deviation of RO estimates in Hill
  curve in Fig.~\ref{fig-sim-hill} for different regions.  Each cell shows the
  bias / standard deviation in percent of the true RO value.  The last columns
  reports the R-squared of the fit, averaged over noise realizations.}
\centering
\begin{tabular}{llllllll}
  \hline
  \multirow{2}{*}{Region} &
  \multirow{2}{*}{Method} & \multicolumn{5}{c}{Concentration (mg/mL)} &
  \multirow{2}{*}{R-squared} \\
  & & 10.0 (RO=2\%) & 100.0 (RO=16.7\%) & 1000.0 (RO=66.7\%) & 10000.0 (RO=95.2\%) & 100000.0 (RO=99.5\%) & \\
  \hline
  \multirow{4}{*}{Striatum}
  & IPR      & 65.5 / 362.0   & 20.0 / 38.9 & -0.1 / 5.8 & -0.4 / 4.9 & -0.6 / 4.7 & 0.981 \\
  & IPRs     & 2.5 / 383.8    & 12.4 / 36.0 & 0.1 / 9.3  & -0.6 / 9.9 & 0.1 / 6.8  & 0.963 \\
  & DPR      & -114.3 / 242.7 & -4.3 / 23.7 & -0.7 / 4.9 & -0.7 / 3.8 & -1.4 / 3.2 & 0.993 \\
  & JDPR     & -85.4 / 343.8  & -1.4 / 33.6 & -5.8 / 7.9 & -5.1 / 6.8 & -5.5 / 6.4 & 0.972 \\
  \hline
  \multirow{4}{*}{Temporal lobe}
  & IPR      & 3303.0 / 523.8  & 298.7 / 61.5 & -0.4 / 14.1  & -28.6 / 11.0 & -33.3 / 8.3   & -32.952 \\
  & IPRs     & 1118.6 / 1142.9 & 80.4 / 104.1 & -42.0 / 31.6 & -63.4 / 19.1 & -57.6 / 21.3  & -3.524  \\
  & DPR      & 2046.3 / 707.6  & 186.6 / 87.5 & -26.5 / 23.9 & -42.1 / 15.9 & -44.7 / 16.6  & -9.668  \\
  & JDPR     & -150.7 / 447.7  & -7.8 / 49.1  & -12.4 / 12.7 & -12.4 / 12.2 & -11.8 / 9.3   & 0.901  \\
  \hline
  \multirow{4}{*}{Whole brain}
  & IPR      & 2575.5 / 206.9 & 233.0 / 24.6 & -4.4 / 4.9   & -29.7 / 3.8  & -32.1 / 3.4  & -24.525 \\
  & IPRs     & 715.2 / 662.2  & 31.3 / 69.3  & -36.2 / 16.5 & -46.9 / 11.9 & -45.7 / 11.4 & -1.872  \\
  & DPR      & 1817.4 / 556.3 & 156.6 / 76.7 & -23.2 / 21.6 & -38.3 / 14.3 & -37.7 / 13.2 & -7.678  \\
  & JDPR     & -146.1 / 434.8 & -7.4 / 47.5  & -11.7 / 12.4 & -11.9 / 11.5 & -11.2 / 9.0  & 0.91    \\
\end{tabular}
\end{table}
\end{document}